\documentclass{article}

\makeatletter
\def\ps@IEEEtitlepagestyle{%
  \def\@oddhead{\mycopyrightnotice}%
}
\def\mycopyrightnotice{%
  \begin{minipage}{\textwidth}
  \scriptsize
   This paper is accepted for publication in \textit{Policy Insights from the Behavioral and Brain Sciences}. Full citation:\\ Nanayakkara, P. \& Hullman, J. What to Consider When Considering Differential Privacy for Policy, \textit{Policy Insights from the Behavioral and Brain Sciences}, 0(0). Copyright © 2024. DOI: https://doi.org/10.1177/23727322241278687.
  \end{minipage}
}
\makeatother

\usepackage{arxiv}

\usepackage[utf8]{inputenc} % allow utf-8 input
\usepackage[T1]{fontenc}    % use 8-bit T1 fonts
\PassOptionsToPackage{hyphens}{url} % wrap urls
\usepackage{hyperref}       % hyperlinks
\usepackage[hyphenbreaks]{breakurl}
\usepackage{url}            % simple URL typesetting
\usepackage{booktabs}       % professional-quality tables
\usepackage{amsfonts}       % blackboard math symbols
\usepackage{nicefrac}       % compact symbols for 1/2, etc.
\usepackage{microtype}      % microtypography
\usepackage{lipsum}
\usepackage{graphicx}
\graphicspath{ {./images/} }
\usepackage{xcolor}
\usepackage{changepage} % for adjwidth
\usepackage{setspace}
\usepackage[backend=biber, uniquename=false, maxbibnames=3]{biblatex}

\usepackage{booktabs}
\usepackage[normalem]{ulem}
\useunder{\uline}{\ul}{}

\title{What to Consider When Considering\\ Differential Privacy for Policy}

\author{
Priyanka Nanayakkara\\
Northwestern University\thanks{The author is currently at the Center for Research on Computation and Society at Harvard University, but performed work for this paper while at Northwestern University.}\\
  \texttt{priyankan@u.northwestern.edu} \\
   \And
Jessica Hullman \\
Northwestern University\\
  \texttt{jhullman@northwestern.edu} \\
}

\begin{document}

\mycopyrightnotice

\maketitle

\begin{abstract}
Differential privacy (DP) is a mathematical definition of privacy that can be widely applied when publishing data. DP has been recognized as a potential means of adhering to various privacy-related legal requirements. However, it can be difficult to reason about whether DP may be appropriate for a given context due to tensions that arise when it is brought from theory into practice. To aid policymaking around privacy concerns, we identify three categories of challenges to understanding DP along with associated questions that policymakers can ask about the potential deployment context to anticipate its impacts.
\end{abstract}

% keywords can be removed

\section*{Tweet}
Differential privacy is a state-of-the-art framework for releasing data summaries while controlling privacy risk to individuals. But prior policy adoptions have led to controversy for reasons not discussed in the lit. We outline key sociotechnical considerations for policymakers.

\section*{Key Points}

\begin{itemize}
    \item Differential privacy is a state-of-the-art framework for protecting individuals’ privacy when releasing data summaries. Under differential privacy, statistical noise is added to data tabulations to obscure individual people’s information while maintaining high-level patterns in the data.

    \item Differential privacy can be applied in a wide range of contexts from a technical perspective. However, adopting differential privacy for policy has led to controversy. Assessing how appropriate it is for a given situation requires considering a host of sociotechnical considerations.

    \item We identify three challenges to reasoning about differential privacy: first, it is naturally at odds with some common understandings of privacy. Second, it can be difficult to estimate its impact on downstream factors like utility of and trust in the data. Third, differential privacy deployments require implementation and analysis workflows likely to be unfamiliar to many practitioners, calling for strategic planning around how to minimize disruption.

    \item We provide a series of questions policymakers can ask about a use case to help overcome challenges in reasoning about differential privacy. Our aim is for these questions to also help policymakers promote privacy protection, knowledge production, transparency, and trust when making decisions about differential privacy.
\end{itemize}

\section{Introduction}
For the 2020 census, the U.S. Census Bureau updated their technical approach to maintaining confidentiality of responses to be based on differential privacy (DP)~\cite{dwork2006calibrating}, a state-of-the-art framework based on a mathematical definition of privacy protection. This change was philosophical as well as technical: rather than advocating a specific \textit{process} for obfuscating individuals’ data, DP defines a \textit{standard} for functions (i.e., algorithms) applied to data. For concreteness, imagine an algorithm that takes in census counts to compute population totals by race in a geographic region. At a high level, the algorithm is differentially-private if we expect its outputs to be similar regardless of whether any particular individual's information is included in the input.

Algorithms often inject statistical noise into analyses to achieve DP. Random noise is drawn from a specified probability distribution (e.g., Laplace or Gaussian~\cite{dwork2014algorithmic}) controlled by a privacy loss parameter, $\epsilon$, also called the ``privacy loss budget.'' The privacy loss budget controls a tradeoff between the strength of privacy protections and accuracy of estimates: smaller privacy loss budgets mean we expect to add more noise, so privacy protection is stronger, but accuracy is lower.

Despite DP's mathematical rigor, the Census Bureau's adoption of DP was met with backlash and debate~\cite{nanayakkara2022s}, both in the popular media and scholarly writing. Many have questioned its necessity given its implications for downstream analysis of census data. One might question whether DP should be applied in other policy contexts. Technically speaking, DP can be applied in a wide range of contexts, but may be a poor fit in light of various social considerations. For example, the protections it offers—which are rigorous, but narrow—may not align with the privacy needs of data subjects (those who contribute their information) in a particular setting. Or, costs it imposes in terms of training analysts to adapt their workflows to constraints imposed by DP may not outweigh its benefits. At the same time, DP has been advocated for policy: in addition to being used by the Census Bureau, who is bound by Title 13 confidentiality requirements, researchers have suggested DP's promise for adhering to requirements outlined in the Health Insurance Portability and Accountability Act (HIPAA)~\cite{ficek2021differential} and the General Data Protection Regulation (GDPR)~\cite{cummings2018role}. Most recently in the U.S. context, a White House executive order\footnote{\url{https://www.whitehouse.gov/briefing-room/presidential-actions/2023/10/30/executive-order-on-the-safe-secure-and-trustworthy-development-and-use-of-artificial-intelligence/}.} named DP as a potential means of advancing trustworthy artificial intelligence.

Unfortunately, many of the sociotechnical trade-offs that policymakers must grapple with when considering DP remain underdiscussed, in part because of DP's newness as a practical intervention. In this paper, we discuss contextual considerations relevant to decisions about adopting DP. Our discussion is organized into three categories of cognitive and sociocultural challenges arising from DP's abstract and mathematical nature, drawing on observations of past DP deployments. First, given the myriad ways privacy can be conceptualized, DP is naturally at odds with some common understandings of privacy. Second, while a goal of privacy-related policy decisions is often to balance privacy needs against the value of data for downstream analysis and decisions, it can be difficult to estimate the impact DP will have on subsequent data use as well as associated trust (in the data among data users, and in the data collection process among data subjects). Third, DP deployments require implementation and analysis workflows that are likely to be unfamiliar to many practitioners, calling for strategic planning around how to minimize disruption.

We provide a series of questions policymakers can ask about specific contexts to help overcome aforementioned challenges and guide them toward contextually-appropriate recommendations about the use of DP. Our recommendations are intended for policymakers deciding whether to implement DP where previously no technical privacy protections existed or to determine whether to replace an existing approach with DP. We summarize our advice and open questions in Table~\ref{tab:summary}.

\begin{table}[]
\begin{tabular}{@{}p{.75in} p{3in} p{3in}@{}}
\toprule
\textbf{Value}                & \textbf{Advice}                                                                                                                                                                                                                                                                                                                                                                                                                                             & \textbf{Open Questions}                                                                                                                                                                                                                                                                                                                                                                                                                                                                                                         \\ \midrule
\textbf{Privacy Protection}   & \begin{tabular}[c]{@{}p{3in}@{}}Consider whether the definition of DP aligns with privacy needs in a particular context. Be aware that DP might conflict with common attitudes toward binary privacy protection, interest in absolute risk, and expected rather than worst-case outcomes.\\ \\ Be aware that the nature and strength of privacy protections offered under DP can vary wildly depending on implementation choices.\end{tabular}                   & \begin{tabular}[c]{@{}p{3in}@{}}What are methods for quantifying absolute disclosure risk (as opposed to relative disclosure risk)?\\ \\ How might we anticipate downstream harms of disclosure stemming from a data privacy violation?\\ \\ How exactly do violations to common DP assumptions alter resulting privacy protections?\end{tabular}                                                                                                                                                                                    \\ \midrule
\textbf{Knowledge Production} & \begin{tabular}[c]{@{}p{3in}@{}}Anticipate negative impacts of DP for knowledge production by simulating the impact of added statistical noise (at different magnitudes) prior to deployment.\\ \\ Anticipate slowdowns in the process of knowledge generation if DP is introduced in place of no or other protection measures. Allocate resources for training data users to cope with the additional noise.\end{tabular}                                       & How can we elicit information about possible future uses of data to anticipate the utility under DP?                                                                                                                                                                                                                                                                                                                                                                                                                            \\ \midrule
\textbf{Trust}                & \begin{tabular}[c]{@{}p{3in}@{}}The addition of statistical noise in published data may cause data users (and others) to lose trust in the data, especially if prior privacy protection measures were kept secret.\\ \\ \\ The use of DP may increase trust in the data collection process among some data subjects.\end{tabular}                                                                                                                                & \begin{tabular}[c]{@{}p{3in}@{}}To what extent does DP impact trust among data subjects and data users relative to no protection or alternative approaches, if at all?\\ \\ What is the relationship between trust and willingness to share data? To what extent are data subjects more willing to share their data when DP is used in a given context?\\ \\ How might we foster participatory engagement around DP deployments to include a wide range of perspectives and strengthen trust in how data are protected?\end{tabular} \\ \midrule
\textbf{Transparency}         & \begin{tabular}[c]{@{}p{3in}@{}}Explore opportunities for making deployment decisions public, auditable, and interpretable beyond the organization deploying DP.\\ \\ \\ If DP is used, provide clear and interpretable explanations of DP's guarantees and impacts to data subjects.\\ \\ \\ If DP is used, provide documentation about the addition of statistical noise and provide guidance for accounting for noise in analysis to data users.\end{tabular} & \begin{tabular}[c]{@{}p{3in}@{}}How should DP guarantees be explained to a lay audience?\\ \\ What types of DP documentation do data users find most helpful for their analysis workflows?\end{tabular}                                                                                                                                                                                                                                                                                                                              \\ \bottomrule
\end{tabular}
\caption{We summarize our advice for promoting four key values—privacy protection, knowledge production, trust, and transparency—when considering DP. DP and its impacts are active areas of research, hence we also include open questions corresponding with each value that, if answered, will further aid policymakers.}
\label{tab:summary}
\end{table}
\section{What Is Differential Privacy and How Is It Different From Prior Approaches?}

Consider a table of published statistics, such as population counts by race and age group. Aggregating data (e.g., by only publishing statistics over groups defined by race and age) may seem to prevent answering questions about any particular individual, but theoretical computer scientists have proven that any informative statistic necessarily leaks information about the people it describes.\footnote{See discussion of the Fundamental Law of Information Recovery in \textcite{dwork2014algorithmic}.} DP provides a framework for accounting ``privacy loss'' incurred with the publication of any given statistic. Specifically, DP is a mathematical standard for the behavior of an algorithm applied to data. It stands in stark contrast to prior approaches. For example, for the 2010 census, the Census Bureau swapped data from carefully-chosen households~\cite{christ2022differential}. $k$-anonymity~\cite{samarati1998generalizing}, another technique, stipulates that published tables must have at least $k$ individuals that share each combination of potentially-revealing attributes (e.g., age and race). Though these approaches are simpler to explain and implement, they are vulnerable to attacks that DP protects against. For example, an adversary can identify whether an individual contributed their data in $k$-anonymous data if they have relevant background knowledge~\cite{machanavajjhala2007diversity}, whereas DP's guarantees are designed to hold regardless of an adversary's prior knowledge (and knowledge they may obtain in the future). We consider an adversary to be anyone who might attempt to deduce individual people's information from published data.

As alluded to earlier, the definition of DP\footnote{The formal definition of DP~\cite{dwork2006calibrating} follows. A randomized algorithm $\mathcal{A}$ is $\epsilon$-differentially-private if for any pair of neighboring databases $D$ and $D'$ that differ by one record, and for any $O \subseteq \mathcal{O}$, where $\mathcal{O}$ is the set of all possible outputs, it satisfies: $$\textup{Pr}\left[\mathcal{A}(D)\in O\right] \leq e^{\epsilon} \textup{Pr}\left[\mathcal{A}(D')\in O\right]$$} describes how similar we expect an algorithm's outputs to be with the addition or removal of any given person's information. Algorithms satisfying DP are randomized. Specifically, DP ensures that the chances of returning any particular output on input datasets that differ by one person's information are close. We refer the interested reader to \textcite{wood2018differential} for an in-depth primer on DP.
\section{Differential Privacy Is at Odds With Common Understandings of Privacy}

DP equates privacy protection with statistically limiting what can be learned about an individual based on analyses of datasets that include the individual~\cite{dwork2006calibrating}. However, privacy is a vast, social concept~\cite{schoeman1984philosophical} that is difficult to fully capture mathematically~\cite{seeman2024between}. We summarize four ways in which common societal understandings of privacy are at odds with DP.

\subsection*{DP Provides Non-Binary Protections}
DP provides \textit{probabilistic} guarantees, yet privacy is often conceptualized as binary: one's privacy is protected or not. Using DP means providing partial privacy protection. The mathematical reality that publishing statistics leaks some information about individuals introduces a question of \textit{how much} privacy loss to permit, which may seem inappropriate to those expecting binary protection.

\textit{Is it required, legally or otherwise, for the data to be published?}

Where data must be collected and released, DP offers a path toward satisfying these needs while providing some privacy protections. Hence, DP is most suited for cases where data collection and publication is required. For example, the Census Bureau must satisfy a dual mandate of both publishing statistics (13 U.S.C. § 141) and maintaining confidentiality (13 U.S.C § 9). DP provides a framework for navigating what appear to be conflicting goals. When there are no requirements to collect data, policymakers should be wary of DP being used to justify invasive data gathering~\cite{sarathy2022algorithmic}.

\subsection*{DP Is Suited to Quantifying Relative Disclosure Risk}
DP is suited to quantifying relative disclosure risk: how much more likely it is that an adversary learns a person's sensitive attribute with access to published data relative to without access (i.e., an adversary is $x$\% more likely to correctly guess Alice's race with the data versus without). However, many find it more natural to use absolute disclosure risk for decision-making: the chance of an adversary learning a sensitive attribute with the published data without comparison to any other value (i.e., the adversary has $x$\% chance of correctly guessing Alice's race with the data). Researchers have argued that absolute disclosure risk is the more relevant measure because relative risk can hide important information~\cite{hotz2022balancing}; for example, a large relative increase in risk may not be cause for major concern if absolute risk is very low to begin with. See \textcite{reiter2005estimating} and \textcite{mcclure2012differential} as cited in \textcite{hotz2022balancing} and \textcite{jarmin2023depth} for further discussion and specific methods of computing disclosure risk.

\textit{To what extent is sensitive information likely to be available to adversaries prior to a data release?}

To contextualize relative risk measures, policymakers should attempt to estimate an adversary's belief about any given person's sensitive attribute prior to seeing the published data. For instance, health insurance companies aiming to learn people's personal information in order to adjust rates may already have access to estimates of the information via data brokers, companies that collect, buy, and sell data about people.\footnote{\url{https://www.propublica.org/article/health-insurers-are-vacuuming-up-details-about-you-and-it-could-raise-your-rates}.} If adversaries can likely access high-quality estimates of the sensitive attribute using data sources other than the one being protected, the application of DP may seem like applying a bandage to a larger problem, where the marginal benefit it provides is lessened. That said, the availability of external information may change over time, for example as the result of legal reductions in data brokers’ permissions. Hence, applying DP may yield benefits in the long run. Policymakers should consider both short- and long-term benefits of using DP.

\subsection*{DP Provides Worst-Case Guarantees}
DP provides worst-case privacy guarantees by upper-bounding how much we expect the results of an analysis to differ based on the inclusion of any given individual's data. For example, using DP with $\epsilon=0.1$ for computing population totals might dictate that at most, the chance of a particular count outputted when Alice is represented in the data is 1.1 ($e^\epsilon$) times the chance of that particular count outputted when Alice is not represented in the data. DP's guarantees hold even assuming an adversary with unlimited computational power and separate datasets to the one being protected. This is why DP is described as aligning with the assumption that potential adversaries will have an increasing amount of both over time.
The notion of making decisions so as to minimize one's maximum loss is familiar in economics, where it is known as the minimax decision strategy. However, often it is preferable to make decisions under uncertainty by choosing an action that is a best response to the most likely outcome~\cite{vonneumann1944theory, savage1954foundation}.

\textit{Is worst-case reasoning about events natural in this context?}

In contexts where the consequences of an event are immense, worst-case reasoning is natural. These are cases in which the consequence of an event is so great that its very possibility is enough to change how decisions are made~\cite{clarke2008possibilistic}, for instance when providing witness or whistle-blower protections.
If the strongest guarantees are unnecessary, but other considerations motivate using DP, it may be useful to consider relaxations of the ``pure'' DP definition ($\epsilon$-DP), such as ($\epsilon$, $\delta$)-DP~\cite{dwork2014algorithmic}. This definition allows for a ``probability of failure'' (i.e., $\delta$), representing the chance that the DP guarantee will not hold. Such relaxations often yield gains in accuracy.

\subsection*{DP Does Not Prevent Making Statistical Inferences}
DP provides protections against threats like (1) linkage attacks~\cite{dwork2014algorithmic}, where an adversary links data without identifiers, like names, to data with identifiers to re-identify people and learn their sensitive attributes, (2) differencing attacks~\cite{dwork2014algorithmic}, where an adversary computes statistics over subsets of the data to learn a person's sensitive attribute, and (3) membership inference attacks~\cite{shokri2017membership}, where the adversary's goal is to determine whether a person's data is included in a dataset. DP does not prevent making statistical inferences~\cite{bun2021statistical}---that is, learning overall patterns about a population---which can lead to apparent privacy violations that rely on making high-confidence guesses about a person's sensitive attribute based on aggregated information~\cite{kenny2021use}. For example, if a statistic indicates that the majority of people living in a particular geographic region belong to a certain racial group, an adversary may use this information to guess with high confidence that Alice, an individual living in that region, belongs to the majority racial group. Entirely restricting inferences of this kind would prevent most scientific knowledge production~\cite{bun2021statistical}. However, to those new to DP guarantees, the fact that DP does not prevent inference can seem contradictory. Indeed, conflicts around DP's protections as related to inference were a point of debate in the Census Bureau's DP deployment for the 2020 census~\cite{nanayakkara2022s}.

\textit{Which, if any, threat models are most concerning?}

Policymakers should consider which threat models—imagined processes by which adversaries could perform attacks—are most concerning. Threat models grounded in real-world concerns can be elucidated through qualitative interviews with data subjects or threat modeling exercises where people are asked to think through potential adversaries and their goals, optionally prompting to consider threats that may disproportionately impact marginalized and vulnerable populations~\cite{sim2023scalable}. These threat models can be compared to those DP protects against to help determine DP's appropriateness. If a combination of attacks are of concern, only some of which DP protects against, policymakers may consider using DP alongside other privacy-enhancing technologies or policies around data use to fill gaps.

\section{Real-World Impacts of Differential Privacy Are Hard to Estimate in Advance}

DP offers an elegant theoretical framework for quantifying the privacy–accuracy tradeoff, but other real-world factors implicated in the tradeoff are more difficult to estimate. We summarize such real-world impacts and offer questions for policymakers that can help them account for these factors when determining DP's appropriateness.

\subsection*{Utility of Data Published Under DP Can Be Difficult to Foresee}

In comparison to accuracy, it is much harder to estimate how DP will impact the utility of data. Utility refers to the usefulness of data beyond simply what is captured by quantitative measures of accuracy, such as the usefulness of the statistic for informing policy~\cite{cummings2024comment}. Utility is difficult to measure because downstream uses of data are potentially vast. Many other analyses may be unknown when the data are published~\cite{hotz2022balancing}. For example, census data are collected for well-defined purposes like reapportionment, but are also used for social science research broadly. Observational data, which are not collected for a particular purpose, may become useful when used in combination with experimental data in a causal inference context~\cite{mann2023combining}.

\textit{To what extent are future uses of the data known and clearly defined?}

When all uses of the data are known and well-defined, DP can be an appropriate choice. For example, think of data that are collected for a single specific purpose, such as to determine the efficacy of a specific policy intervention. In such cases, policymakers can commission simulations of how noise introduced by DP will likely impact outcomes. Prior work can serve as a useful guide to commissioning such studies. For example, researchers have empirically investigated how DP may impact decisions driven by U.S. census data, such as the allocation of federal funds, voting rights benefits, and redistricting~\cite{pujol2020fair, steed2022policy, kenny2021use, cohen2022census} and how DP might impact estimates of small-population geographies and minority populations~\cite{kenny2024evaluating, christ2022differential}. Policymakers may find it helpful to follow or adapt practices from prior work, like testing a range of settings for the DP deployment (such as through different values of the privacy loss budget) or comparing the magnitude of DP noise to other sources of error to better contextualize DP's impact~\cite{steed2022policy, kenny2024evaluating}.

Policymakers should solicit information about what kinds of analyses data users plan to conduct, and run simulations accordingly. Policymakers may consider developing a set of ``acceptance criteria''---attributes about the published data that should be maintained after DP is applied---alongside data users and other relevant parties to ensure utility downstream, as was done with a recent differentially-private synthetic data release from Israel's National Registry of Live Births~\cite{hod2024differentially}. Finally, policymakers should be aware that estimating downstream utility is a challenge they may face with other approaches as well, and is not unique to DP.

\subsection*{How DP Impacts Trust in Data Collection and Published Data is Not Well Understood}

Using DP may increase trust in data collection among data subjects, potentially leading to higher response rates. On the other hand, the explicit addition of statistical noise introduced by DP may rupture a ``statistical imaginary'' around data being ``objective and neutral''~\cite{boyd2022differential}, leading to a decrease in trust in the fidelity of published data among the broader public and data users. This decrease may be especially salient if data users are used to seeing the data without privacy protection or with forms of protection that are necessarily less visible to uphold the method's integrity (known in computer science as ``security through obscurity''~\cite{kerckhoffs_principle}). In the case of DP applied to the U.S. census—which was previously protected through “invisible” methods like swapping—the addition of DP noise brought attention to the ways in which census data are not “a raw headcount of the public that the Census Bureau simply collects and tabulates”~\cite{boyd2022differential}. We hold the view espoused by \textcite{manski2020lure} that encouraging a veneer of ``incredible certitude'' is ultimately counterproductive to policy and data literacy. Policymakers should nevertheless anticipate how DP may abruptly collide with existing ``statistical imaginaries''~\cite{boyd2022differential}.

\textit{What resources exist to ensure data users and data subjects will be effectively informed about the use of DP?}

Policymakers should assess whether there are adequate resources to invest in communication around the integrity of the published data. This communication should include how DP noise and other sources of noise (e.g., undercounting in a census context) impact the data. Furthermore, policymakers should invest in fostering recruitment and training of ``trusted local experts''~\cite{abdu2024algorithmic} (in this case, experts within groups of data users with knowledge of data users’ practices, needs, and attitudes as well as DP) to ensure that communication artifacts are effectively received by data users. Policymakers can also learn from sources of conflict in previous deployments of DP to proactively address potential miscommunications~\cite{nanayakkara2022s} and from past efforts to increase transparency around data releases under DP (e.g.,~\cite{abdu2024algorithmic, hod2024differentially}).

Similarly, a growing body of work aims to explain DP guarantees to data subjects (e.g., ~\cite{nanayakkara2023w, xiong2020towards}), with some findings suggesting that explanations of DP may improve trust and increase willingness to share data. In scenarios where data subjects are concerned about past data collection and publication efforts, understandable explanations of privacy measures that are possible under DP may ease their concerns and foster trust in data collection. A risk is, of course, that tensions between DP and common notions of privacy challenge widespread understanding of DP's guarantees.

\subsection*{The Impacts of Violations to Common DP Assumptions Are Not Yet Well Understood}

Accounting privacy loss under DP relies on assumptions which are often at odds with needs to keep certain statistics unaffected by noise. For example, state population counts from the 2020 census were held invariant—meaning no statistical noise was added—for legal reasons related to apportionment.\footnote{\url{https://www2.census.gov/library/publications/decennial/2020/census-briefs/c2020br-04.pdf}.} The architects of the census DP deployment have noted the unknown privacy impacts of invariants~\cite{garfinkel2018issues}; in fact, their impact is the topic of ongoing research~\cite{bailie2023can}. Beyond theoretical measures of privacy loss, empirical measures can also be useful in understanding the extent to which such violations impact privacy (e.g.,~\cite{kenny2021use}), but such work remains relatively limited and inherently context-specific.

\textit{Will the deployment require straying from conventional DP approaches?}
Policymakers should determine whether the data they plan to publish under DP have a straightforward structure that will not require bespoke differentially-private algorithms. For example, in the case of the U.S. census, the Census Bureau had to invent and implement a new algorithm to accommodate various specificities of census data, including its hierarchical nature, where smaller geographic units comprise larger geographic units~\cite{abowd2022census, garfinkel2018issues}. Hence, DP may be adopted with less financial costs and more quickly in cases where invariants and other violations are not anticipated.

\subsection*{Harms of Disclosure May Be Difficult to Estimate}
Real-world harms are harder to estimate than disclosure risk. For example, if an adversary discovers Alice's sensitive attribute, it is difficult to predict with high confidence how they will use this information and what impact it will have on Alice or others. Publicized examples of major harms resulting from leaked data provide partial information about the landscape of risks, such as identity theft.\footnote{\url{https://www.usatoday.com/story/money/2024/01/25/data-breach-id-theft-protection/72352690007/}.} Recent legal scholarship proposes a typology of privacy harms, which include physical, economic, and psychological harms, among others~\cite{citron2022privacy}. To account for these possibilities in decision-making, policymakers must be ready to associate them with costs, broadly defined, so they can be formally reasoned about.

\textit{Why do people hesitate to share their information?}

A natural approach to identifying possible harms of disclosure is to understand why people hesitate to share their information. Engaging with data subjects can take several forms, including qualitative interviews, surveys, or focus groups. For example, the Census Bureau has historically conducted surveys and interviews to learn why people hesitate to respond to the census~\cite{mcgeeney20192020}, and external researchers have similarly been able to identify reasons for not participating in the census by talking to people~\cite{NAP25978}. Investigations of this nature may surface concerns that fall outside the sole purview of technical solutions, like government collusion.

\subsection*{How to Effectively Facilitate Participatory Engagement Is Not Yet Well Understood}

The decision of whether to adopt DP and under what specifications may benefit from participatory engagement, considering DP impacts a range of parties like data subjects and data users. However, how to foster productive engagement is not obvious. Prior deployments, like that of the 2020 U.S. Census, have demonstrated that different parties arrive at discussions with varied assumptions and epistemic orientations around data~\cite{boyd2022differential}, which make participatory engagement challenging at best, and a liability for progress at worst~\cite{eyal2019crisis}.

\textit{What are people's epistemic commitments and orientations to the data, and how can these be accounted for in participatory processes?}

Unfortunately, there is no one-size-fits-all guidance for facilitating participatory engagement well in the context of DP deployments. Interventions must be developed to account for the fact that in each use case, relevant parties likely have different epistemic commitments and orientations to the data. For example, data users may view access to data differently in a scenario where DP is making the release of data possible versus one where the data were previously released without any technical protection. How to account for varied perspectives during discussions about DP is an ongoing, and critical area of future work.

\section{Deploying Differential Privacy Requires New Analysis Practices and Skills}

Applying DP entails making a series of implementation choices. These choices are likely unfamiliar to data curators. Moreover, the statistical noise introduced by DP will change analysts’ typical routines. We present three implementation-related challenges, along with avenues to ensuring smoother DP deployments, both before and after data publication.

\subsection*{Applying DP Entails Navigating Complex Implementation Choices}
Unfortunately, there is little existing guidance for curators making DP implementation choices. For example, DP experts largely consider setting the privacy loss budget to be a ``policy question''~\cite{garfinkel2018issues}. However, this parameter is unit-less and does not directly map to real-world outcomes, making it difficult to reason about even among policy experts. Curators must also make a series of other choices. For example: the “unit” at which protections apply (e.g., the record level or the user level)~\cite{dwork2019differential}; which variation of DP to use; and the deployment model (such as the central model, where raw data are collected and noise added by the curator, or the local model, where data are noised prior to being sent to the curator~\cite{kasiviswanathan2011can}). Making these choices requires weighing various tradeoffs. For example, the local model affords additional privacy protections—because data subjects never share raw information—but comes at the cost of lower accuracy than the central model. Research to develop interactive interfaces for curators (e.g.,~\cite{nanayakkara2022visualizing}), which would help them make these decisions, is still early stage.

\textit{How can policy requirements guide curators toward making prudent deployment decisions?}

Policymakers can consider establishing limits on parameter ranges data curators must work within. However, doing so will likely require significant research. Alternatively, they can recommend curators follow guidance from reputed sources, such as NIST~\cite{near2023guidelines}.\footnote{While NIST's guidelines~\cite{near2023guidelines} are primarily intended to \textit{evaluate} DP guarantees, they are also a valuable reference for decisions made pre-deployment.} Or, they may advise the adoption of a framework that blends DP with philosophical frameworks~\cite{benthall2024integrating} to help integrate normative reasoning into otherwise abstract and technical decisions.

\subsection*{Depending on How DP is Implemented, the Privacy Guarantees Can Vary Wildly}
Implementation choices can have profound impacts on strength of privacy protections. Thus, DP can be used to create the appearance of privacy without substantive protections—in other words, ``privacy theater''~\cite{smart2022understanding, seeman2024between, sarathy2022algorithmic}.

\textit{Are there existing infrastructure and practices to enact auditing or otherwise make implementation choices public? If not, can they be put in place?}

Competing incentives around privacy protection can manifest in organizational settings in various ways, such as emphasis on ``fitness for use'' of data over privacy~\cite{steed2024adoption}. Hence, policymakers should ensure systems of accountability and transparency prior to a deployment. For example, an ``Epsilon Registry'' could serve as a public repository of implementation choices across DP deployments~\cite{dwork2019differential}. Policymakers may require or strongly recommend that organizations deploying DP make all implementation choices public.

These choices should ideally be presented in a format understandable by people with and without technical DP expertise. To enable audiences with technical expertise to perform audits, policymakers may further recommend that source code for the deployment is made public. Making implementation choices public prior to deployment can also enable an organization's competitors or other relevant parties to push back against ill-advised plans. Such was the case when Google's competitors pushed back against their plans to cluster users into granular groups, causing Google to adapt their plan~\cite{steed2024adoption}.

\subsection*{Analyzing and Interpreting Data Under DP Calls for New Analysis Practices}
Analysts can be given access to data either via differentially-private algorithms or through published data products with added statistical noise. In both cases, they will have to cope with changes in their routines. Analysts who are given access to the data via differentially-private algorithms are given access to the data under a total privacy loss budget. With each query for a statistic they wish to compute, they must ``spend'' some portion of the budget. Once the budget is depleted, they may no longer issue queries. This constraint will reconfigure their typical practices~\cite{sarathy2023don}, particularly for exploratory analyses, since they can only make a limited number of queries. Hence, they will need to efficiently allocate their budget across queries while still learning what they need to about the data; research is just beginning to explore how to support these needs (e.g.,~\cite{nanayakkara2024measure}). On the other hand, analysts who are given access to DP-noised data products will need to account for how DP noise impacts uncertainty around their findings~\cite{hotz2022balancing}. This may sometimes require the use of new statistical methods. Such changes should be expected to introduce hiccups in workflows and possibly pushback.

\textit{Who will be given access to the data via differentially-private algorithms, and what is their training? How can analysts be supported in interpreting data noised under DP?}

If data are made available through differentially-private algorithms, policymakers should consider the statistical backgrounds of potential analysts. The transition to DP can be managed by offering this audience statistical training that will equip them to handle new practices, like spending a privacy loss budget.
Analysts who work with DP-noised data products should also adapt their routines. Policymakers should consider how to communicate the new layer of noise in ways that are interpretable and usable, including documentation that explains the added noise and how to account for it in analyses. For instance, the data may contain seemingly nonsensical values owing to DP noise, like negative counts, causing confusion if not explained well~\cite{garfinkel2018issues, dankar2013practicing}. Policymakers should also consider how to set up feedback loops between analysts and the research community, who is developing techniques for analyzing DP-noised data products and accounting for the additional layer of statistical uncertainty. Interdisciplinary workshops with both data users and researchers can promote feedback loops and help ensure that methods are developed for common use cases.\footnote{As an example, see a recent workshop on analysis of DP-noised census data: \url{http://dimacs.rutgers.edu/events/details?eID=2038}.}
\section{Conclusion}
Challenges to understanding DP may prevent policymakers from making informed recommendations about using DP. However, policymakers can more easily anticipate potential impacts and make recommendations that incorporate lessons from prior DP deployments by asking context-specific questions. Moreover, as technical methods of protecting privacy grow in availability and sophistication, guides like ours may support policymakers in making technical privacy recommendations more broadly.

\section*{Acknowledgments}
We thank danah boyd, Christian Cianfarani, Ryan Steed, and participants of the Privacy Law Scholars Conference---especially Jayshree Sarathy, our discussant---for their thoughtful feedback on an earlier draft of this work.

\section*{Declaration of Conflicting Interests}
The authors declared no potential conflicts of interest with respect to the research, authorship, and/or publication of this article.

\section*{Funding}
The authors received no financial support for the research, authorship, and/or publication of this article.

\printbibliography

\end{document}